\documentclass[11pt,twoside]{article}

\usepackage{asp2014}

\aspSuppressVolSlug
\resetcounters

\begin{document}

\title{Shall numerical astrophysics step into the era of Exascale computing?}

\author{Giuliano~Taffoni,$^1$ Giuseppe~Murante,$^1$ Luca~Tornatore,$^1$ David~Goz,$^1$ Stefano~Borgani,$^1$ Manolis~Katevenis,$^2$ Nikolaos~Chrysos,$^2$ and Manolis~Marazakis,$^2$
\affil{$^1$INAF Osservatorio Astronomico di Trieste, Trieste, Italy; \email{taffoni@oats.inaf.it}}
\affil{$^2$Foundation for Research and Technology, Heraklion, Crete, Greece}}

% This section is for ADS Processing.  There must be one line per author.
\paperauthor{Giuliano~Taffoni}{taffoni@oats.inaf.it}{0000-0002-4211-6816}{INAF}{OATs}{Trieste}{ }{34131}{Italy}
\paperauthor{Giuseppe~Murante}{giuseppe.murante@inaf.it}{0000-0002-5155-130X}{INAF}{OATs}{Trieste}{ }{34131}{Italy}
\paperauthor{Luca~Tornatore}{tornatore@oats.inaf.it}{}{INAF}{OATs}{Trieste}{ }{34131}{Italy}
\paperauthor{Manolis~Katevenis}{kateveni@ics.forth.gr}{}{FORTH}{}{Heraklion}{Crete}{GR  700 13}{Greece}
\paperauthor{Nikolaos~Chrysos}{nchrysos@ics.forth.gr}{}{FORTH}{}{Heraklion}{Crete}{GR  700 13}{Greece}
\paperauthor{Manolis~Marazakis}{maraz@ics.forth.gr}{0000-0002-4768-3289}{FORTH}{}{Heraklion}{Crete}{GR  700 13}{Greece}

\begin{abstract}

\noindent
High performance computing numerical simulations are today one of the more effective instruments to implement and study
new theoretical models, and they are mandatory during the preparatory phase and operational phase of any scientific experiment. 
New challenges in Cosmology and Astrophysics will
require a large number of new extremely  computationally intensive  simulations to investigate physical  processes at different scales.
Moreover, the  size and complexity of the new generation of observational facilities also implies a new generation of high performance data reduction and analysis tools pushing toward the use of Exascale computing capabilities.
 Exascale supercomputers cannot be produced today.  We  discuss the major technological challenges in the design, development and
 use of such computing capabilities and we will report on the progresses that  has been made in the last years in Europe, in particular in the framework of the  ExaNeSt European funded project. We also discuss the impact of this new computing resources on the numerical codes in Astronomy and Astrophysics.
\end{abstract}

\section{Introduction}
The last decade has seen the advent of  numerous digital sky surveys across a range of wavelengths (e.g the Cosmic 
Microwave Background (CMB) experiments \citep{planck,wmap}, or the Sloan Digital Sky Survey \citep{sloan}, with  terabytes 
of data and often with tens of measured parameters associated  to each observed object. Moreover, during the next decade,
new highly complex and massively large data sets are expected by novel and more complex scientific 
instruments  that might  provide new insights  on the knowledge of
fundamental laws of physics at cosmological scales and in on the formation and evolution of cosmic structures
%the Astronomical and Astrophysical (A\&A) community 
(e.g. the Square Kilometer Array (SKA)  \citep{ska}, 
the Cherenkov Telescope Array  (CTA) \citep{cta}, the Extremely Large Telescope E-ELT \citep{eelt}, the James Webb Space telescope \citep{jwst}, etc).

At the same time, a  new generation of large surveys,  such as the ground based Large Synoptic Survey Telescope \citep{lsst} (LSST), 
or the  Euclid satellite missions  \citep{euclid}  might
shed light on the nature of dark energy and dark matter  and possibly revolutionize modern physics. 

Handling and exploring these new data volumes, and actually making real scientific discoveries, poses a considerable
technical challenge that requires the adoption of new approaches for {\em write} needs to overcome the traditional research methods
and requires the access to a new generation of computing facilities and computational HPC algorithms.

In Astronomy and Astrophysics (A\&A),  High Performance Computing (HPC) numerical simulations are today one of the more effective instrument
to compare observations with theoretical models.
They enable Astronomers to understand nuanced predictions, as well as shape experiments more efficiently.  
They are mandatory during the preparatory and operational phases of new scientific experiments.
They also help capture and analyze the torrent of experimental data being produced by the new generation of scientific
instruments. Computational modeling can illuminate the subtleties of complex theoretical models, 
making HPC infrastructures a theoretical laboratories to test physical processes.

The more accurate these theoretical experiments are,
the more efficient the future large scale surveys will be in solving the mysteries of our Universe, so
the size and  complexity of the new experiments require extremely large computational resources, 
pushing toward  the use of Exascale computing capabilities.

The development of Exascale computing facilities with machines capable of executing O(10$^{18}$) floating point operations per second
(FLOPS) will be characterized by significant and dramatic changes in computing hardware architecture from current  
petascale capable super-computers.  
To build  an Exascale resource we need to address some major technology challenges related to Energy consumption, 
Network topology, Memory and Storage, Resilience and of course Programming model and Systems software. 

From a computational science point of view, the architectural design of existing peta-scale supercomputers, 
where computing power is mainly delivered by accelerators (GPU, FPGA, Cell processors etc.), 
already impacts on scientific applications. 
This will become more evident on the future Exascale resources that will involve millions of processing units causing 
parallel application scalability issues due to sequential application parts, 
synchronizing communication and other bottlenecks. 
Future applications must be designed to make systems with this number of computing units efficiently exploitable.

An approach based on hardware/software (HW/SW) co-design is crucial to enable Exascale computing by solving the
application-architecture performance gap (the gap between the peace capabilities of the HW and the 
performance released by HPC SW) and contributing to the design of supercomputing resources that can 
be effectively exploited by  real scientific applications.

This paper will summarize the major challenges to Exascale  and how much progress has been made in the last years in Europe. 
We will present the effort done by the ExaNeSt EU funded project to build  a prototype of an Exascale facility based on 
ARM CPUs and accelerators, designed using a HW/SW co-design approach, where Astrophysical codes are 
playing a central role in defining network topology and storage system.  
Finally we will discuss how the co-design will impact on Numerical Codes that must be re-engineered to profit of the Exascale supercomputers.

\section{Technical challenges in  Exascale computing}
A new generation of supercomputer  able to 
perform 10$^{18}$ FLOPs,  is expected to be available for the end of 2020.  
They will be between ten and one hundred 
times faster that actual Tier-0 HPC facilities 
(as listed in the TOP500 (http://www.top5000.org) rank). 

\articlefigure[width=0.8\textwidth]{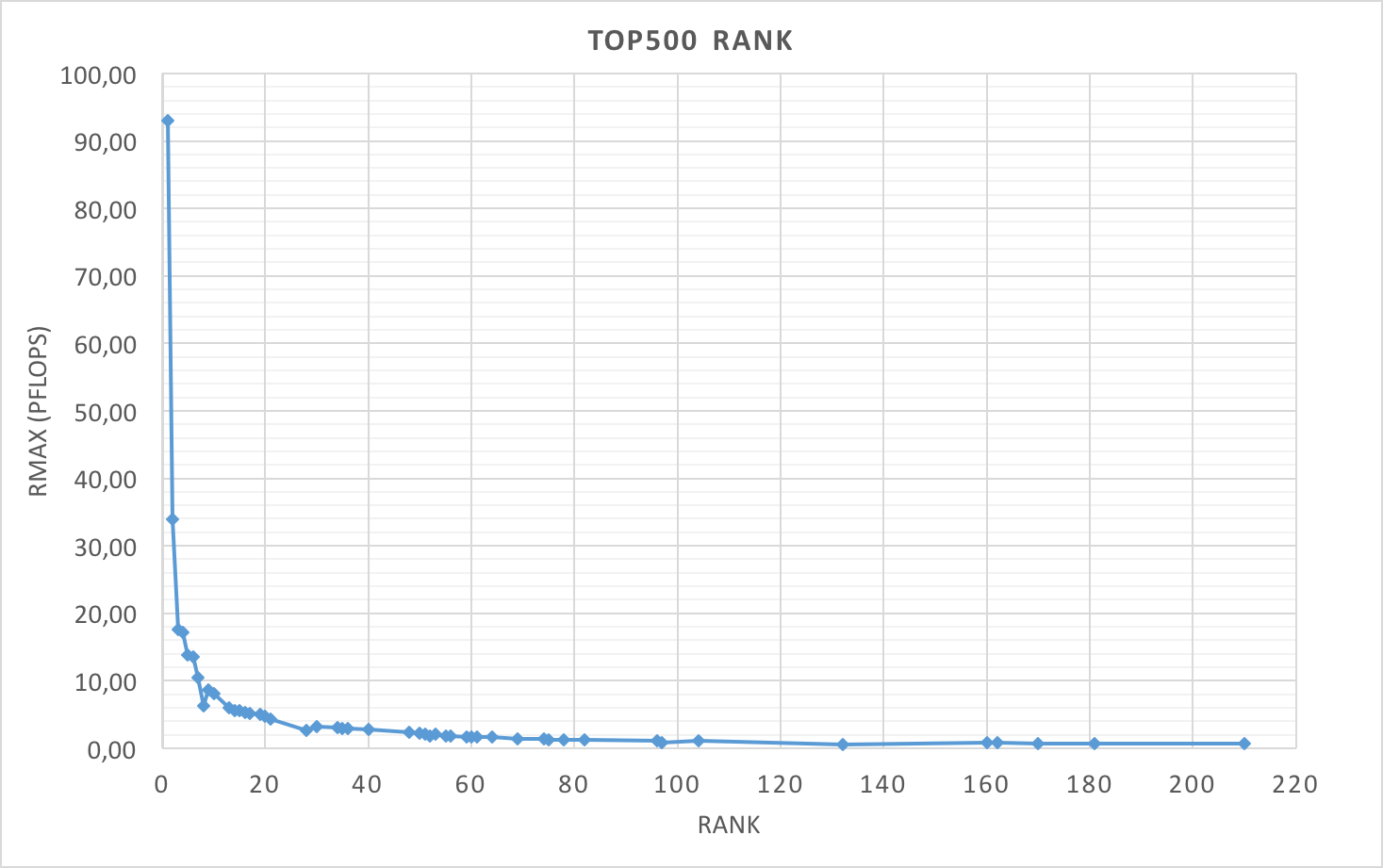}{fig1}{Performance measurement in PFLOPs for the first 200 most 
powerful supercomputers in the World. Performance are  measured with the HPL (matrix operations) package \citep{hpl}.
The first supercomputer in the Top 500 rank is the Chinese Sunway (MPP) that performs 93 PFLOPs  
with $~$10 million cores.}

The realization of an Exascale supercomputer requires significant advances in a variety
of technologies  both HW  and SW.  A series of studies in the last years in  Europe (http://www.etp4hpc.eu), 
USA  \citep{usa} and Japan \citep{Kobayashi2015}
categorize the  technological challenges in few core research topics to face: (a) High performance interconnect technology; (b)  Memory technology (both DRAM and non volatile low-latency high density data storage); (c) System software and runtime system that  ensures that applications are able to maximize the capabilities of the underling HW; (d) Programming systems; (e) Data  management for simulated and observed data; (f) Resilience.

Algorithms and scientific FW improvements contribute to the increasing of the computational capabilities as much as the HW innovation.  Algorithms and codes re-engineering, supported by the proper programming model and system SW is mandatory to exploit the  Exascale platforms.

Beside the technological challenges listed above, there are other two 
crucial aspects that are also deeply related to all of them: the energy efficiency and a new approach towards 
the evaluation of the  computing performance (Sustained performance)
of super computers 

With current semiconductor technologies, all proposed Exascale designs would consume hundreds
of megawatts of power. If we just scale up the actual most powerful supercomputer to reach 
Exascale capacity, it will require about one GW to be operated. New designs
and technologies are needed to reduce this energy requirement to a
more manageable and economically feasible level. And those technologies does not involve only 
the design of the CPUs (that should implement a  simpler but energy efficient design \citep{usa}), 
but also all the  other aspects including algorithms and SW. 
The goal is to  operate and Exascale  facility with less that 50 MW.

\subsection{Sustained Performance versus Peak Performance}
The last decade, the peak performance of high-end computing systems is incredibly boosted by aggregating a
huge number of nodes, each of which consists of multiple fine-grain cores, because
the LINPACK benchmark \citep{hpl} used to rank facilities in the   top 500,  is only computation-intensive.

\articlefigure[width=0.8\textwidth]{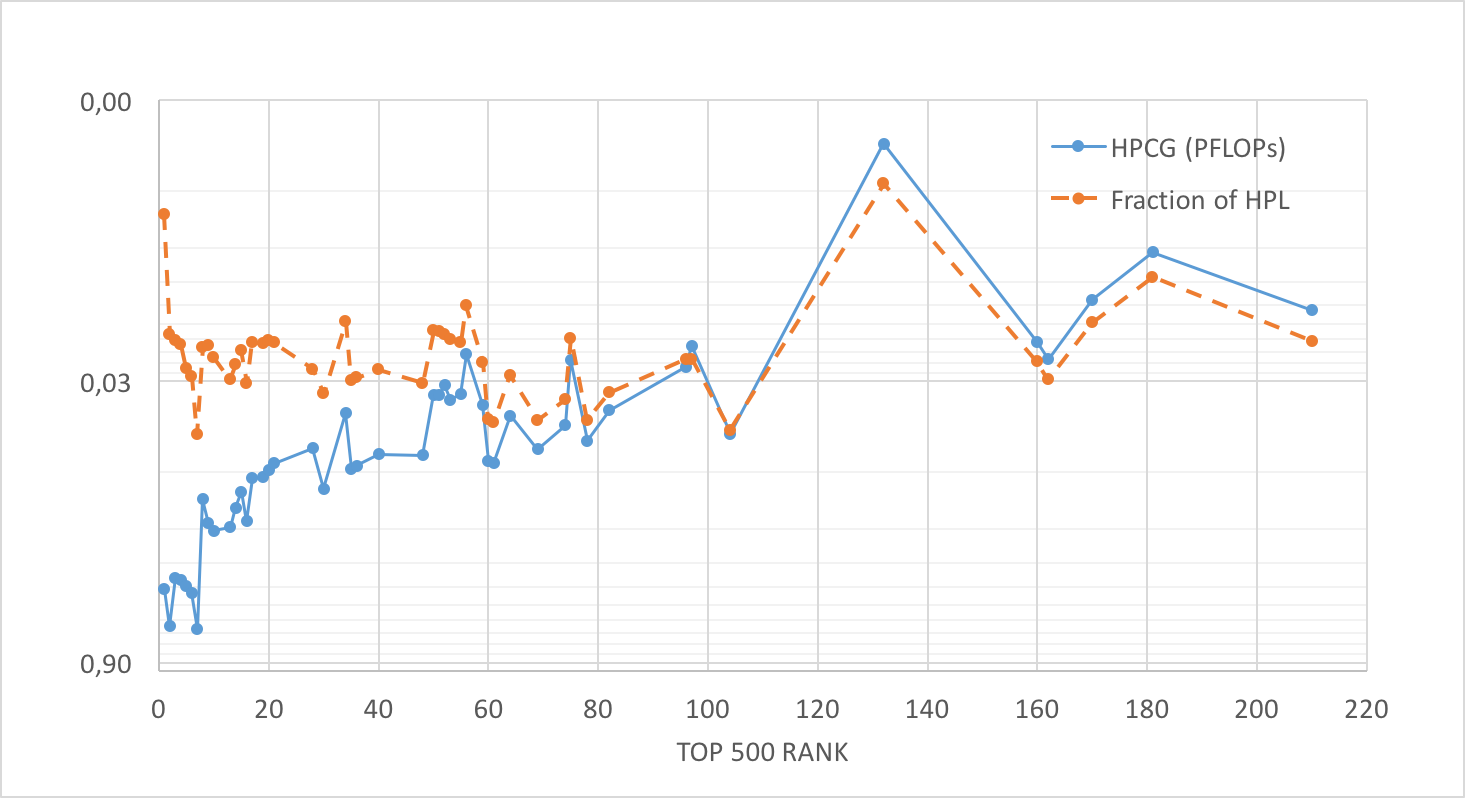}{fig2}{HPCG PFLOPs for the first 200 most powerful supercomputers in the World.
We present the sustained performance of the first  200 most powerful supercomputers in the World as measured with HPCG. 
As reference we plot also the ration between the HPCG and HPL performance.  Sustained performance of the  Chinese Sunway is
10$^{-3}$ less then the HPL measured performance.}

However, many scientific  applications, including N-Body Hydrodynamic Cosmological simulations, are  memory-intensive: they are not only consuming floating point operations but they also need to access
data in memory.

We measure the application behavior in terms of  bytes per flop (B/F), that can be defined as a ratio of a memory
throughput in bytes/s to a computing performance in flop/s of a HPC system.
A recent paper from \citep{Kobayashi2015} show that commonly scientific applications need 0.5B/F or more in the case of Astrophysical applications ($>1 B/F$).

\citet{hpcg} develop a new benchmark tool  to account for sustained performance: the HPCG.

As show in fig.~\ref{fig2}, the sustained performance of actual supercomputers is not even the PFLOPS, and 
scaling up to Exascale in sustained performance means a jump of more that 4 order of magnitude. 

Moreover, applications that implements a high ($~$1) B/F, impact also on the power consumption. For example, today an extremely low power, high frequency RAM   GDDR5 consume about  4.3 W / 64 Gbs
To feed a modest 0.2 B/F for a sustained $10^{18}$ FLOPS, it is  required 
about 12MW of power only for memory access.

\section{The ExaNeSt approach to Exascale}
With architecture evolution, the HPC market has undergone a
fundamental paradigm shift. The adoption of low-cost, Linux-based
clusters extended HPC's reach from its roots in modeling and
simulation of complex physical systems to a broader range of
industries, ranging from cloud computing and deep learning, to automotive and
energy, many of which were originally served by datacenters. 
ExaNeSt \citep{exanest} belongs to a group of ongoing EU projects supporting the next
step forward in this direction. Today, low-power microprocessors dominate the embedded, smartphone
and tablets markets, outnumbering x86 devices both in volume and in
growth rate. If these trends continue, we can expect to see
such energy-efficient servers benefiting from the same economies of scale that in the
past favored general-purpose personal computers (over mainframes) and more
recently commodity clusters (over custom supercomputers).

ARM is the industry leader in power-efficient processor design. ARM
processors consume about 2 to 3 times less electric energy for a given
amount of computation relative to Intel-based processors, and are
widely used in embedded consumer electronics, including mobile phones
and tablets. As a result, many research and industry programs see
ARM-based energy-efficient servers as a potential successor of x86 and POWER-based
servers in hyperscale datacenters supercomputers.

ExaNeSt partners with a number of concurrent FET-HPC projects aim to
collectively answer the HPC challenges described in the strategic
vision statement of ETP4HPC. Common across all projects
is the technological approach for scalable, low-power and economically
viable solution for compute, as is being refined and realized in the
EuroServer \citep{euroserver}. EuroServer has common participants
across ExaNeSt and various other consortia. Here is a summary of the
projects that are aligned with this approach:
\begin{itemize}
\item ExaNeSt \citep{exanest}: is responsible for the physical
  deployment characteristics to support the required compute density,
  along with the storage and interconnect services.
\item ExaNoDe (http://www.exanode.eu): focuses on the delivery of low-power
  compute elements for HPC.
\item ECOSCALE \citep{ecoscale}: focuses on integrating and exposing the
  acceleration capabilities of FPGA's in HPC.
\item Eurolab-4-HPC (http://www.eurolab4hpc.eu): is a supporting action CSA that is to support the actions and initiatives around delivery strategy for ETP4HPC vision.
\end{itemize}

Together, these and other initiatives, at local and international level, have the goal to deliver a European solution for high
performance computing and to ensure Europe's leadership in HPC.

\subsection{The co-design approach for software and hardware}
An intensive co-design effort is essential to build an Exascale system. 
The design of the ExaNeSt platform is tailored to real scientific and industrial applications
used to  define the requirements for the architecture and, at later stage, 
to evaluate the final solution. The behavior of scientific applications is analyzed to 
optimize the  interconnect and the data management of the platform both in terms of HW and SW. 
Applications has been instrumented to  provide their network and I/O traces and the
traces has been analyzed to understand the type and kind of message exchanged by application
tasks (Fig~\ref{traces}).

On the other side applications will be re-engineered to exploit the ExaNeSt platform,  
parallel algorithms will be adapt to it, and new algorithms and implementations will 
be developed to approach the computational capabilities of the new co-designed HW.

\subsection{ExaNeSt Technology: challenges and solutions}
Presently, ExaNeSt is designing two prototype systems based on
Iceotope's packaging technology (http://www.datacenterknowledge.com/archives/2013/03/04/iceotope-liquid-cooling-in-action/). The overall objective is to stress 
our complete solution (interconnect, storage, systems software) using real-world HPC applications, 
which will be ported on the prototypes. The systems that are developed will based on
Xilinx Ultrascale+ FPGAs, with quad ARM Cortex-A53 64-bit cores. Each
compute daughter-board will consist of four FPGAs (i.e. with a total of 16 cores) and
an NVMe in-node SSD storage device.

The first prototype (Track-1) will exploit existing liquid immersion
technology from Iceotope, capable to cool the thermal drive of 800W
per blade. Each blade in Track-1 will host 4-8 compute
daughter-boards (16-32 FPGAs). The total size of Track-1
prototype will depend on the cost of Xilinx FPGAs: with our current price estimates, 
it is expected to range between 6 and 16 blades. Track-2 prototype will use novel 
liquid cooling, developed by Iceotope during course of the ExaNeSt project. 
The new cooling technology will enable even denser designs, with 16 compute daughter-boards per blade.

\subsubsection{Rack-level shared memory}

The ExaNeSt project develops architectures for systems with
densely-packed compute, memory and storage devices, interconnected
using high-performance interconnects. For many big data and HPC
applications, the access to fast DRAM memory is a key to performance.

In our design, the memory attached to each compute node has modest
size --i.e., tens of GB per compute node; in order to make many
hundreds of DRAM available to each compute node, we will plan to
enable remote memory sharing.

Our memory architecture will be based on {\it Unimem}, first developed
within EuroServer \citep{euroserver}. With Unimem, a node can access
parts of memories located in remote nodes. To eliminate the complexity
and the costs of system-level coherence protocols
\citep{laudon1997sgi}, the Unimem architecture defines that each
physical memory page can be cached at only one location. In principle,
the node that caches a page can be the page owner (the node with
direct access to the memory device) or any other remote node; however,
in practice, it's preferred that remote nodes do not cache page.

In ExaNeSt, we are extending Unimem to operate efficiently on a large
installation with real applications. One important enhancement is to enable 
a {\it virtual} global address space, rather
than a physical one, as was the case in Euroserver.
This improves security, allows page migration, and can also simplify multi-programming, just
as virtual memory did in the past for single node systems.

\subsubsection{In-node Data Storage}
Modern HPC technology promises ``true-fidelity''
scientific simulation, enabled by the integration of huge sets of data
coming from a variety of sources. As a result, the problem of ``big
data'' in HPC systems is rapidly growing, fueling a shift towards {\it data-centric HPC architectures}. 
Low-power, low-cost, fast non-volatile memories (NVM) (e.g.
flash-based) is a first key enabling technology for data-centric HPC.
The decreasing cost of low-power fast non- volatile memories (NVM)
(e.g. flash-based) is changing dramatically the computing landscape.
These devices promise to narrow the storage-processor performance gap
with low latency (tens of microseconds vs. 10+ milliseconds) and high
I/O operations per second (IOPS) performance 
at capacities of several hundreds GBytes.
ExaneSt will place
these storage devices with the compute nodes rather than in a
centralized location.
We propose extensions to a parallel file system to take advantage of such devices
as cache layer. Moreover, we design cache maintenance protocols based
on the concepts of Unimem memory consistency model. Our high-level
goal is that as long as the processors stay within the (extended)
coverage offered by their local (in-node) storage devices, they
achieve low-latency and low-power access to data, avoiding the
movement of data across long distances.

\subsubsection{Unified Interconnect}
The ExaNeSt project 
focuses on a tight integration of fast NVM devices at the
node level using Unimem to improve on data locality. The presence of
distributed low-latency devices 
introduces new challenges for the underlying interconnect. 
In this project, we advocate the need for a unified system interconnect that
will merge inter-processor traffic with a major part (if not all) of
storage traffic. This consolidation of networks is expected to bring
significant cost and power benefits, as the interconnect is
responsible for 35\% of the power budget in
supercomputers and consumes a lot of power even when it idles
\citep{hoefler2010software}.

The most advanced
inter-processor interconnects, although customized to provide
ultra-low latencies, typically assume benign, synchronized
(application- clocked) processor traffic  
%\citep{denzel2010framework}.
Pulling the bursty storage flows inside a unified interconnect may
negatively affect performance, thus calling for advanced
Quality-of-Service (QoS) support. We will address the design of a
suitable unified interconnect for exascale systems by splitting the
work into two major parts. In the first part, we  address the
interconnect within a rack, examining suitable low- power electrical
and optical technologies and appropriate topologies. We address system packaging and topology selection in tandem,
aiming at multi-tier interconnects \citep{kodi2014photonic}, to address
the disparate needs and requirements at separate building blocks
inside the rack (e.g. mezzanine board, chassis, etc.). A set of
alternative state-of-the-art photonic and optical link technologies
will be explored.

\section{ExaNeSt Applications and Co-design: Astrophysical Numerical simulations}
A set of relevant and ambitious code from different scientific areas has been identified for co-design, including HPC for astrophysics, nuclear physics, neural networks  and big data.

In A\&A, often the scientific problems under investigation involves complex physics, 
a very large dynamical range, or both.  
This kind of computation requires high resolution, that translates 
in a very large number of computational elements - usually particles. 

In the ExaNeSt project we use the TreePM+SPH code GADGET-3, evolution of the public code 
GADGET-2 \citep{Springel2005} and PINOCCHIO semi-analytical code \citep{pinocchio}. 
In simulations used in ExaNeSt, besides gravity and hydrodynamics, the following processes 
are modeled: radiative cooling of the gas, star formation,
chemical evolution, energy feedback from exploding stars,
a uniform time-dependent UV background,
evolution of black holes and energy outputs from them.

This numerical computation are dynamical in space and time:
the computation evolves from a initially homogeneous, easy to balance configuration, 
to an extremely dis-homogeneous one. Furthermore, galaxies move with respect to each other, possibly colliding and clustering together. This behaviour makes the load-balancing a severe issue and those codes an excellent and challenging candidate to design and optimize
the interconnect of a supercomputer. 

In ExaNeSt project, we used two different cosmological simulations. 
The first one is a portion of the Universe, a cube having a side of 25 Mpc, 
with relatively low resolution (details in \citet{Barai2015}). 
This simulation does not resolve the internal structure of galaxies 
and is thus relatively easy to balance in all its phases. 
The second one follows the birth and the evolution of a single galaxy. 
Here, the resolution is higher and the dis-homogeneity towards the 
end of the simulation is larger (details in \citet{Murante2015}). 
In neither case we run the full simulation, which would be very time-consuming. %that is computationally quite expensive. 
So we instrument the code and run the simulation to collect network traces for 
small temporal intervals  at the beginning of the simulation, at intermediate 
times and towards the end,  so as to sample the different dynamical situations, 
that can reflect in different communication patterns and workload balances.
An example of traces collected is shown in Fig.~\ref{traces}.

\articlefigurethree{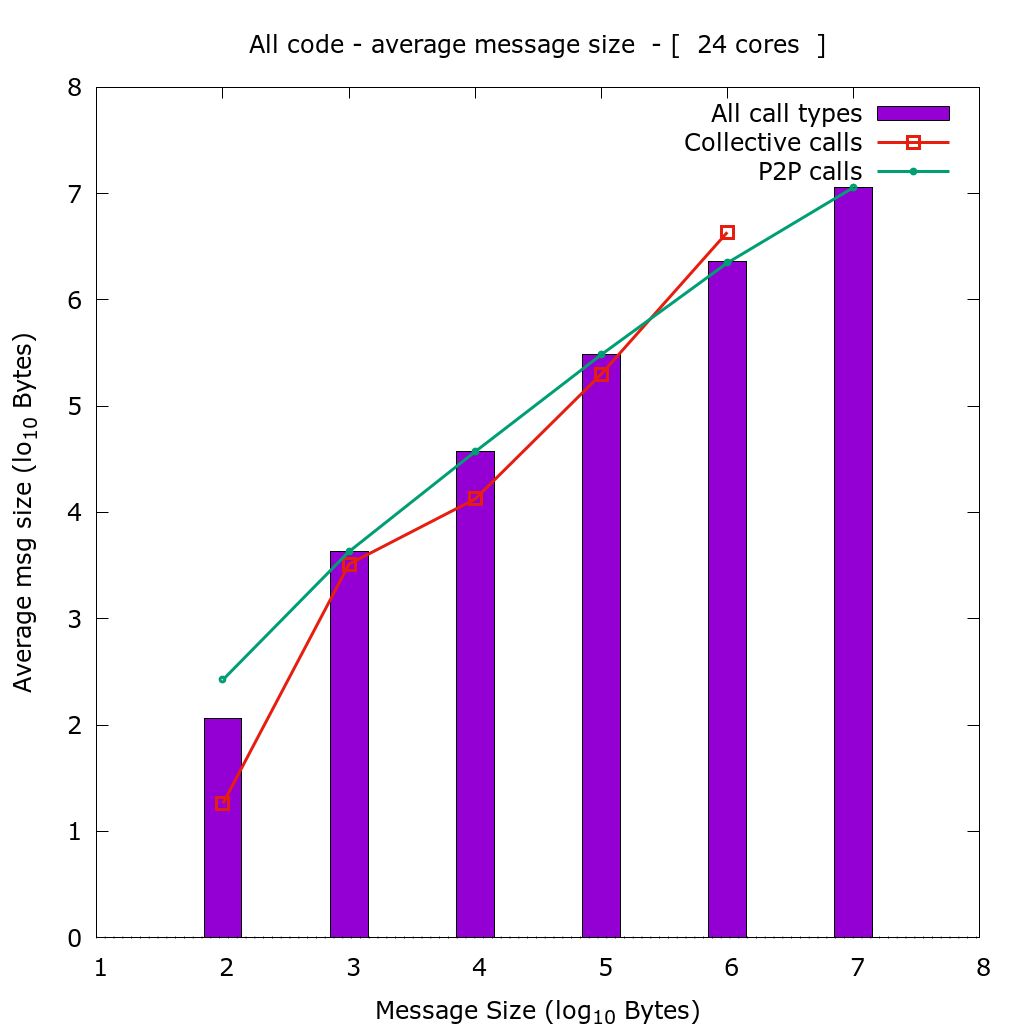}{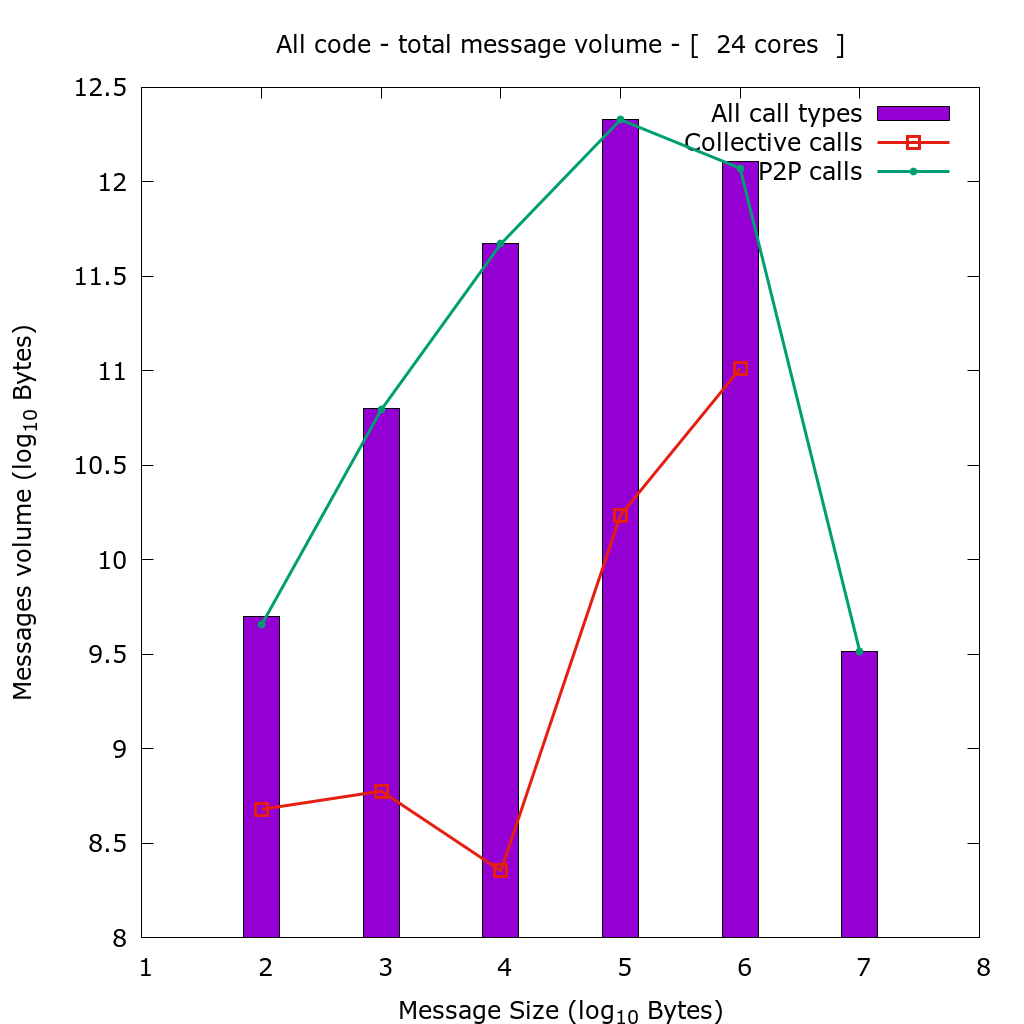}{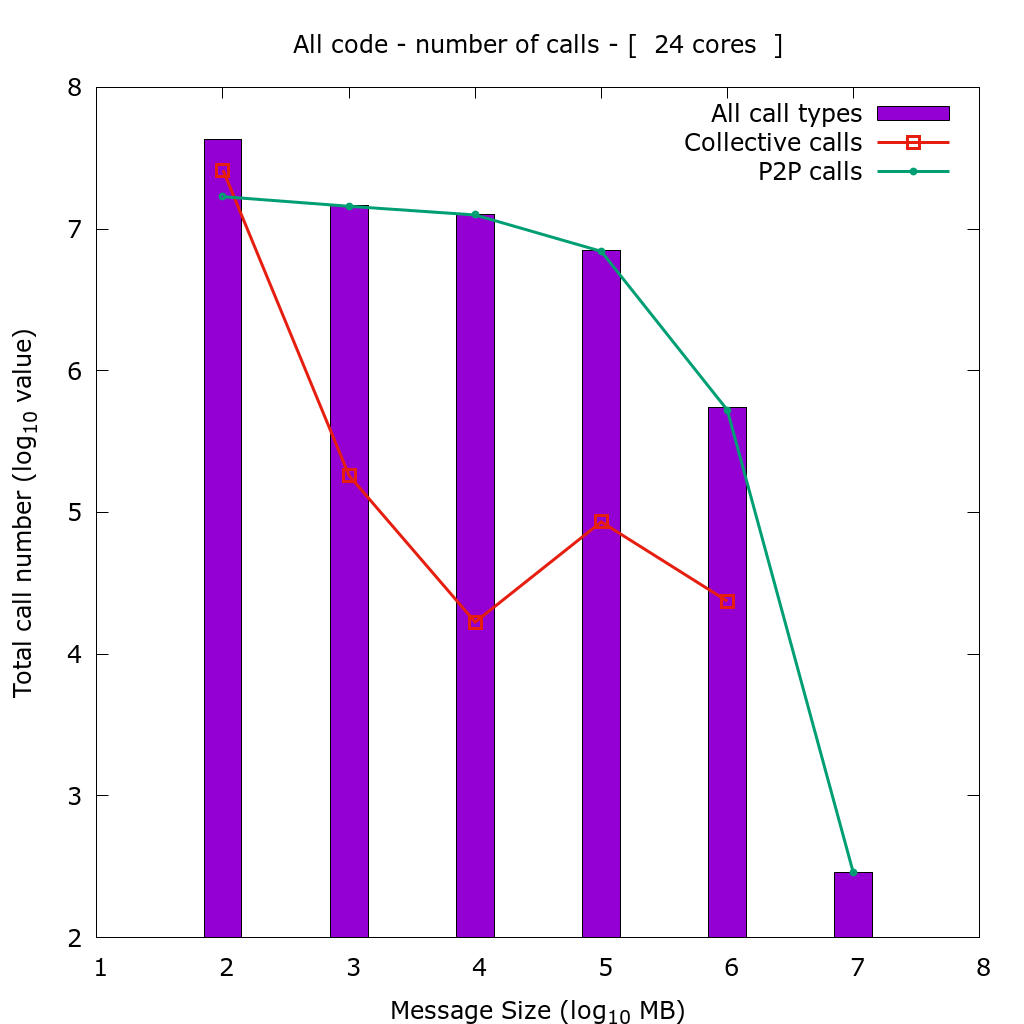}{traces}{The plots show the average message size [\textit{left panel} the total amount of data [\textit{central panel}]], both in $log_{10}$ Bytes, and the total number of calls to MPI\_ functions [\textit{right panel}]. All data have been binned per message size to which they refer to (x axis, $log_{10}$ Bytes).}

\section{Implementing an exascalable Astrophysical HPC application}

GADGET is a full-fledged high-performance code for cosmological simulations. However, some architectural concerns arise at the edge of exascale era.

Parallel machines are increasingly built with (cc)NUMA hierarchical architectures in which multiple cores have access to a hierarchy of memories. Multiple cache
memories are placed at the lowest (fastest) levels of the hierarchy, and every core may
have individual caches, while groups of cores may share higher level caches or single
memory controllers.
On such systems it is crucial that codes take advantage of spatial and temporal locality of data, which also requires to be NUMA-aware. This
concept translates in the "affinity" that a given memory segment has with a particular computing core.
The GADGET memory model is not NUMA-aware and tries to exploit memory locality
only with some basic stratagem.
Hence the re-design of the code must start from the re-design of the memory model so that to consider the different affinity of different memory regions.
Due to the extreme diversity of physical processes being modeled and algorithms implemented, this a difficult task that does not have a unique solution and may require
memory layout transformations in some points.

Secondly, the code has been conceived as the parallel generalization of a serial code, in a pre-multithread era. Hence the workflow is rigidly procedural, meaning by this that all the
tasks perform the same operations - possibly individually using more threads in local
loops- with frequent synchronization via MPI message.
Fig.~\ref{traces} reveals at glance that the number of communications
that exchange very small amount of data (few to hundreds of bytes) largely outnumber
those that exchange large data chunks (0.1-10 or more MB), although the latter account
for most of the total data traffic on the machine network. As a consequence, the network
latency can represent a significant fraction of both the communication and running times.
In view of these two facts, and the strongly hierarchical architecture foreseen for the exascale machines, it is compulsory to adopt a different code design i.e. to decompose the workflow in as-small-as-possible single tasks with clear dependencies on, and conflicts between, each other from both the point of view of operations to be performed and data to be processed.
In such a way, the workflow would be translated in a queue system where idling threads
perform the first available tasks on not-under-use data. Synchronization of operations
should pass as much as possible through RDMA operations and the queue system itself.
Moreover, 'encouraging' threads that resides on the same group of cores to undergo similar tasks, or tasks operating on the same data, should lead to a more efficient exploitation
of memory affinity and locality (even if redundancy of some data may be required).

We will start from a stripped-down version of the publicly available code, that incorporates
only the gravity and hydrodynamical solvers, and the star formation plus the stellar evolution
modules. We plan to separately develop mini-apps for each of the core algorithms, designed
ab-initio as ''atomic'' inter-dependent tasks.
In this way we will be able to develop different algorithms and strategies for each of
the ''pillars'' detailed above in a much easier way that in a unique monolithic code.

\section{Conclusions}
Today supercomputers must be considered as theoretical laboratories 
necessary to Astronomers as much as any observational facilities. 
New generation of super computers will be able to perform ${10^{18}}$ 
capabilities however it is unlikely   that Exascale is achievable
without disruptive changes in the way the super computers will be  
built and in the way we will use them.
Astronomers will be obliged to re-engineer their applications 
terms of a new paradigm  based on task based programming, 
innovative resilience and heterogeneous computing 
(CPUs/GPUs/HW accelerators).
Moreover, it will be necessary to bring computation 
close to data improving. As a consequence, only applications that implement a low B/F will exploit Exascale computations.
This will be in practice extremely complex when dealing with  Big Data, opening new challenges for A\&A community in prevision of the new experiments.

\acknowledgements 
This project has received funding by the European Union's Horizon 2020 Research and Innovation Programme under the ExaNeSt project (Grant Agreement No. 671553).

\bibliography{I12.1}  % For BibTex

\end{document}